\documentstyle[11pt,epsf,epsfig]{article}
\newcommand\C{\v Cerenkov }

\begin{document}

\title{A Water \v Cerenkov Calorimeter as the Next Generation Neutrino Detector}

\author{Yi-Fang Wang \\
Stanford University, Department of Physics \\
Stanford, CA 94305,USA \\
E-mail: yfwang@hep.stanford.edu}


\maketitle

\abstract{We propose here a large homogeneous calorimeter as the
next generation neutrino detector for $\nu$ factories and/or 
conventional $\nu$ beams. The active media is chosen to be water for
obvious economical reasons. The \v Cerenkov light produced in water 
is sufficient to have good energy resolution, and the pattern recognition is 
realized by a modular water tank structure.
Monte Carlo simulations demonstrate that the detector performance is excellent for
identifying neutrino CC events while rejecting background events.
}

\section{Introduction}

Neutrino factories and conventional  beams 
have been discussed
extensively in the literature\cite{nuf}
as the  
main facility of neutrino physics for the next decade. 
The main physics objectives include
the measurements of $sin\theta_{13}$, $\Delta m^2_{13}$, the leptonic
CP phase $\delta$
and the sign of $\Delta m^2_{23}$. 
All of these quantities can be obtained through the 
disappearance probability $\mathrm P(\nu_{\mu}\rightarrow\nu_{\mu})$ and
the appearance probability 
$\mathrm P(\nu_{\mu}(\nu_e)\rightarrow \nu_e(\nu_{\mu}))$ and
$\mathrm P(\bar\nu_{\mu}(\bar\nu_e)\rightarrow \bar\nu_e(\bar\nu_{\mu}))$.
To measure these quantities, a
detector should:
1) be able to identify leptons: e, $\mu$ and if possible $\tau$;
2) have good pattern recognition capabilities for background rejection;
3) have good energy resolution for event selection and to determine 
$\mathrm P_{\alpha\rightarrow\beta}(E)$;
4) be able to measure the charge for $\mu^{\pm}$ in the case of
 $\nu$ factories; and
5) be able to have a large mass(100-1000~kt) 
at an affordable price.

\begin{table}[htbp!]
\begin{center}
\begin{tabular}{|l|c|c|c|c|}
\hline
              & Iron        & Liquid      & Water Ring & Under Water/Ice  \\
              & Calorimeter & Ar TPC      & Imaging    & \v Cerenkov counter \\ 
\hline
Mass          &  10-50 kt        & 1-10 kt     & 50-1000 kt  & 100 Mt   \\
Charge ID     &  Yes             & Yes            & ?          & No       \\
E resolution    & good             & very good    & very good  & poor    \\
Examples      & Minos            & ICANOE       & Super-K, Uno & Amanda, Icecube \\
              & Monolith         &              & Aqua-rich  & Nestor,Antares \\
\hline
\end{tabular}
\end{center}
\caption{Currently proposed detector for $\nu$ factories 
and conventional $\nu$ beams.}
\end{table}

Currently there are four types of detectors proposed\cite{nuf,dick}, as listed
in table~1.
These detectors are either too 
expensive to be very large, or 
too large to have a magnet for charge identification. 
In this talk, I propose a new type of 
detector -- a water 
\C calorimeter -- which fulfills all the above requirements.

\section{Water \C Calorimeter}

Water \C ring image detectors have been successfully employed 
in large scale, for obvious economic reasons,
by the IMB and the Super-Kamiokande experiments.
However a substantial growth in size 
beyond these detectors appears problematic because of 
the cost of excavation and photon detection. 
To overcome these problems, we propose here 
a water \C calorimeter with a modular structure,
as shown in Fig. 1.

\begin{figure}[htbp!!]
\begin{center}
\mbox{\epsfig{file=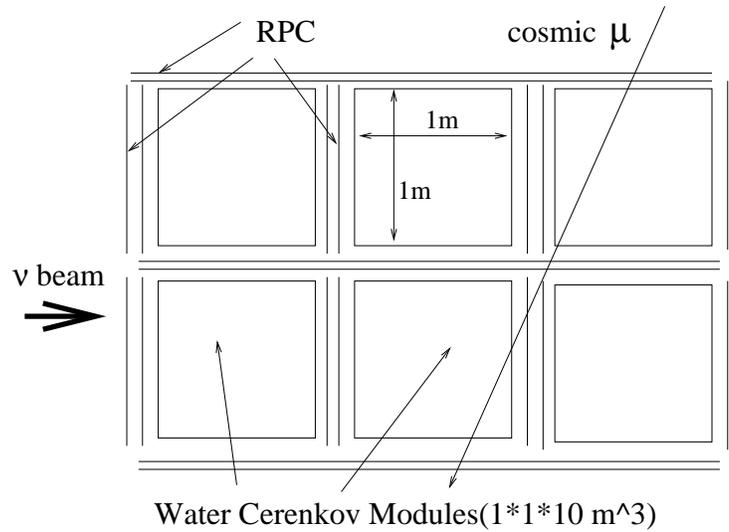,height=7.cm}}
\caption{Schematic of water \C calorimeter }
\end{center}
\end{figure}

\begin{figure}[htbp!]
\begin{center}
\mbox{\epsfig{file=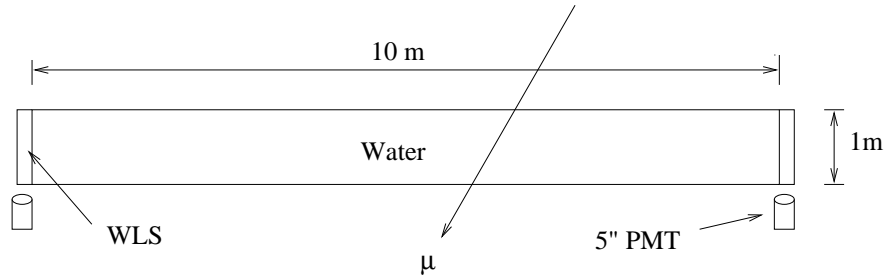,height=3.6cm}}
\caption{Schematic of a water tank.}
\end{center}
\end{figure}

Each tank has  
dimensions $\mathrm 1\times 1\times 10 m^3$, holding a total of 10~t of water.
The exact segmentation of water tanks is to be optimized based on the neutrino beam energy,
the experimental hall, the cost, etc. For simplicity, we discuss in the following
1~m thick tank,
corresponding to 2.77 X$_0$ and 1.5 $\lambda_0$. 
The water tank is made of PVC with Aluminum lining.
\v Cerenkov light is reflected by Aluminum and transported
towards the two ends of the tank, which are covered by 
wavelength shifter(WLS) plates. 
Light from the WLS is guided to a 5'' photon-multiplier tube(PMT), as shown in Fig. 2. 
The modular structure of such a detector allows it to be placed 
at a shallow depth in a cavern of any shape(or possibly even at surface), 
therefore reducing the excavation cost. The photon collection area is also
reduced dramatically, making it possible to build a large detector 
at a moderate cost.

A through-going charged particle emits about 
20,000 \C photons per meter. Assuming a light attenuation length in water of
20m and a reflection coefficient of the Aluminum lining of 90\%, we obtain 
a light collection efficiency of about 20\%. 
Combined with the quantum 
efficiency of the PMT(20\%), the WLS collection efficiency(25\%) and an additional 
safety factor of 50\%, the total light collection efficiency is about 0.5\%.
This corresponds to 100 photoelectrons per meter, 
which can be translated to a resolution of $\mathrm 4.5\%/\sqrt{E}$.
This is slightly worse than the Super-Kamiokande detector and 
liquid Argon TPC but much better than
iron calorimeters\cite{nuf}.

If this detector is built for a $\nu$ factory,  
a tracking device, such as
Resistive Plate Chambers (RPC)\cite{rpc} will be needed between water tanks 
to identify the sign of charge.
RPCs can also be
helpful for pattern recognition, to determine precisely muon directions,
and to identify cosmic-muons for either veto or calibration. 
The RPC strips will run in both X- and Y-directions with a width of 4 cm.
A total of $\sim 10^5 ~m^2$ is needed for a 100 kt detector, which is more than
an order of magnitude larger than the current scale\cite{rpc}.
R\&D efforts would be needed to reduce costs.

The magnet system for such a detector can be segmented in order to minimize dead 
materials between water tanks.
If the desired minimum muon momentum is 5 GeV/c, the magnet must be segmented
every 20 m. Detailed magnet design still needs to be worked out; here we just present 
a preliminary idea to start the discussion.
A toroid magnet similar to that of Minos, as shown in Fig. 3, can produce a 
magnetic field $\mathrm B>1.5~\mathrm T$, for a current $\mathrm I> 10^4$~A.
The thickness of the magnet needed is determined by the error from
the multiple scattering:
$\mathrm \Delta P/P = 0.0136\sqrt{X/X_0}/0.3BL$,
where L is the thickness of magnet.
For L=50 cm, we obtain an error of 32\%.
The measurement error is given by 
$\mathrm \Delta P/P \simeq \delta\alpha/\alpha=
\sigma P/0.3rBL$,
where r is the track length before or after the magnet
and $\sigma$ is the pitch size of the RPC. For P=5 GeV/c, $\sigma=4$ cm and r=10 m,
the measurement error is 9\%, much smaller
than that from multiple scattering. 
It should be noted that $\mathrm P_{\mu}$ is also measured from the range. By requiring
that both $\mathrm P_{\mu}$ measurements are consistent, we can eliminate most of the fake wrong 
sign muons. 
The iron needed for such a magnet is about 20\% of the total mass of the water.

\begin{figure}[htbp!]
\begin{center}
\mbox{\epsfig{file=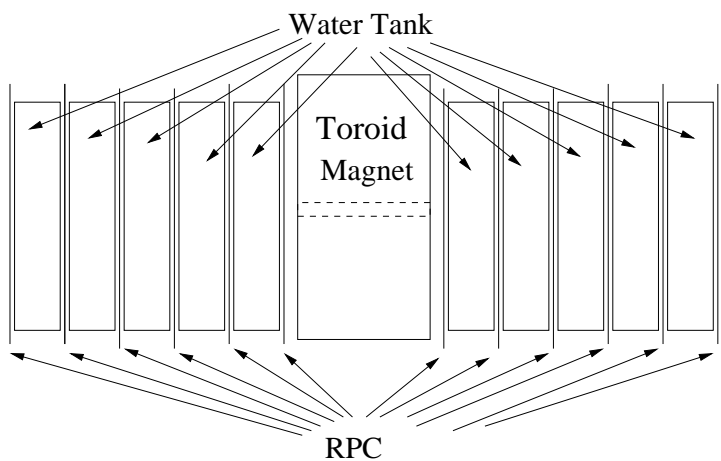,height=6.cm}}
\caption{Schematic of a toroid magnet}
\end{center}
\end{figure}

The cost of such a detector is moderate compared to other types of detectors, enabeling
 us
to build a detector as large as 100 - 1000 kt. 
The combination of size, excellent energy resolution and 
pattern recognition capabilities makes this detector very attractive. 
An incomplete but rich physics program can be 
listed as follows:
1) neutrino physics from $\nu$ factories or $\nu$ beams;
2) improved measurements of atmospheric neutrinos; 
3) observation of supernovae at distances up to hundreds of kpc;
4) determination of primary cosmic-ray composition by measuring multiple muons;
5) searches for WIMP's looking at muons from the core of the earth or the sun with a 
sensitivity covering DAMA's allowed region;
6) searches for monopoles looking at slow moving particles with high dE/dx;
7) searches for muons from point sources;
8) searches for exotic particles such as fractionally charged particles.
Depending on the location of the detector, other topics on 
cosmic-ray physics can be explored.

\section{Performance of Water \v Cerenkov Calorimeter}

To study the performance of such a detector, 
we consider in the following 
two possible applications in the near future: JHF neutrino beam 
to Beijing with a baseline of 2100 km and NuMi beam from Fermilab to 
Minos with a baseline of 735 km. 
The energy spectra of visible $\nu_{\mu}$ CC events are 
shown in Fig.~4.

\begin{figure}[htbp!]
\begin{center}
\mbox{\epsfig{file=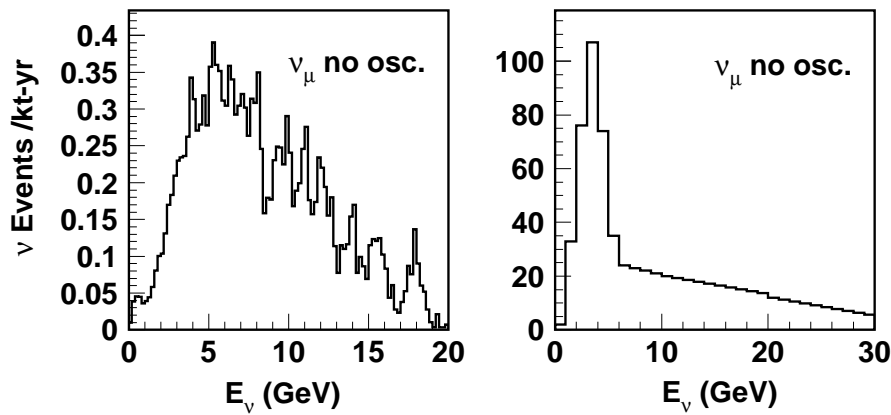,height=6.cm}}
\caption{Beam profile of JHF-Beijing and Numi-Minos.}
\end{center}
\end{figure}

We use a full 
GEANT Monte Carlo simulation program and the Minos neutrino event 
generator. A CC $\nu$ signal event is identified by its 
accompanying lepton, reconstructed as a jet.
Fig. 5 shows the jet energy normalized by the energy of the lepton.
It can be seen from the plot that leptons from CC events 
can indeed be identified and the jet reconstruction algorithm works properly.
It is also shown in the figure that the energy resolution of the neutrino CC events
is about 10\% in both cases.

\begin{figure}[htbp!]
\begin{center}
\mbox{\epsfig{file=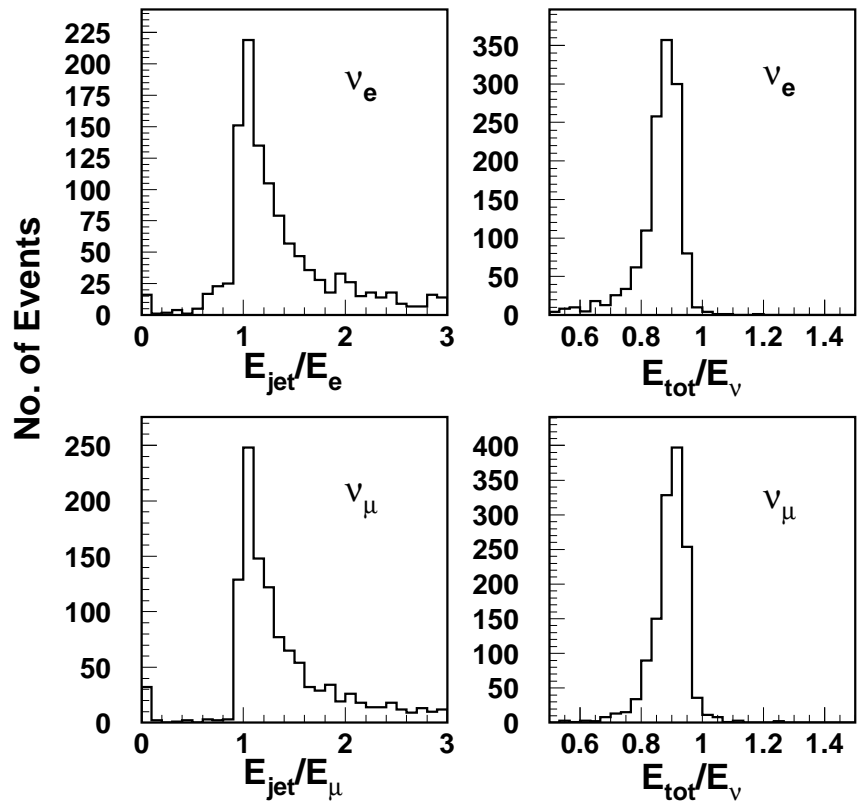,height=11.cm}}
\caption{The reconstructed jet energy and total visible energy. 
The fact that $\mathrm E_{jet}/\mathrm E_{lepton}$
peaks around one shows that the jet reconstruction algorithm 
finds the lepton from CC events. The fraction of total visible energy 
to the neutrino
energy indicates that we have an energy resolution better than 10\% for all neutrinos. 
The bias is due to invisible neutral hadrons and charged particles below \C thresholds.}
\end{center}
\end{figure}

\begin{figure}[htbp!!!]
\begin{center}
\mbox{\epsfig{file=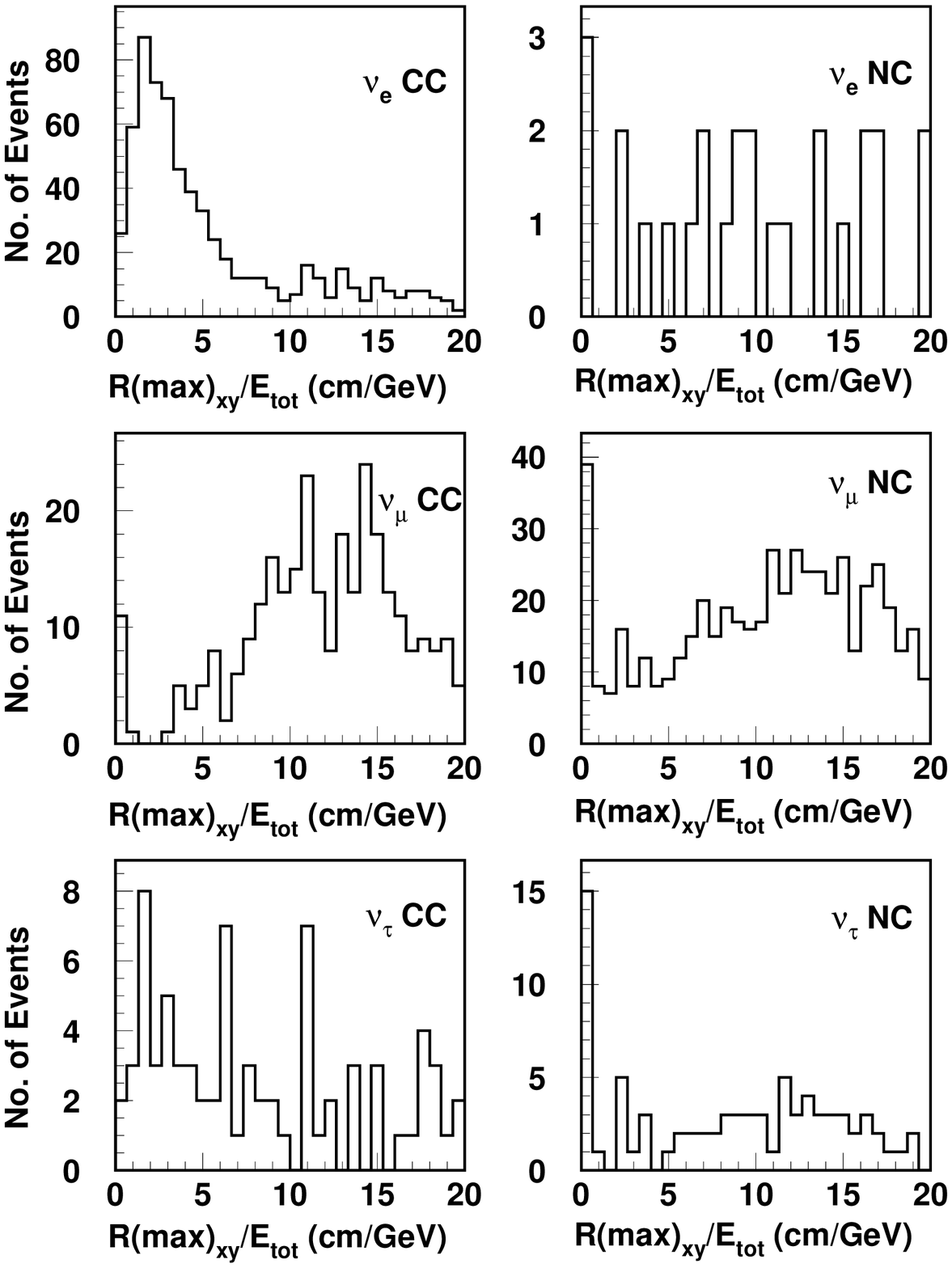,height=16.cm}}
\caption{The transverse event size at the shower maxima for various type of 
neutrino events. The distribution of $\nu_e$ is different from all the others.}
\end{center}
\end{figure}

The neutrino CC events are identified by the following 5 variables:
$\mathrm E_{max}/\mathrm E_{jet}$, $\mathrm L_{shower}/\mathrm E_{jet}$, 
$\mathrm N_{tank}/\mathrm E_{jet}$,
$\mathrm R_{xy}/\mathrm E_{tot}$, 
and $\mathrm R^{max}_{xy}/\mathrm E_{tot}$, where 
$\mathrm E_{jet}$ is the jet energy, $\mathrm E_{tot}$ the total visible energy,
$\mathrm E_{max}$ the maximum energy in a cell,
$\mathrm L_{shower}$ the longitudinal length of the jet,
$\mathrm N_{tank}$ the number of cells with energy more than 10 MeV, 
$\mathrm R_{xy}$ the transverse event size and
$\mathrm R^{max}_{xy}$ the transverse event size at the shower maxima.
Fig. 6 shows $\mathrm R^{max}_{xy}/\mathrm E_{tot}$ for all
different neutrino flavors. It can be seen that $\nu_e$ CC events can be selected 
with reasonable efficiency and moderate backgrounds.
Table~2 shows the final results from this pilot Monte Carlo study.
For $\nu_e$ and $\nu_{\mu}$ 
events, $\nu_{\tau}$ CC events are dominant backgrounds, while for $\nu_{\tau}$,
the main background is $\nu_e$. It is interesting to see that this detector can identify
$\nu_{\tau}$ in a statistical way. Similar results are obtained for a detector
with 0.5m water tanks without RPCs. 
These results are similar to or better than those from water \C image 
detectors\cite{other}
and iron calorimeters\cite{wai2}. 

\begin{table}[htb!]
\begin{center}
\begin{tabular}{|l|c|c|c|c|c|}
\hline
      & \multicolumn{3}{|c|}{JHF-Beijing} & \multicolumn{2}{|c|}{NuMi-Minos} \\
      & $\nu_e$ & $\nu_{\mu}$ & $\nu_{\tau}$ & $\nu_e$ & $\nu_{\mu}$ \\
\hline
CC Eff.             & 30\%  & 53\%      & 9.3\%&  15\% &  53\%  \\
\hline
$\nu_{e}$ CC        &  -    & $>$1300:1 & 3:1   &  -    & $>$1300:1 \\ 
$\nu_{e}$ NC        & 166:1 &  665:1    & 60:1  & 600:1 & $>$610:1 \\
$\nu_{\mu}$ CC      & 700:1 &   -       & 270:1 & 14000:1 &   -    \\
$\nu_{\mu}$ NC      & 92:1  & $>$6000:1 & 39:1  & 320:1   & 2000:1 \\
$\nu_{\tau}$ CC     & 20:1  & 12:1      &  -    & 33:1    & 18:1 \\
$\nu_{\tau}$ NC     & 205:1 & 1100:1    & 61:1  & 530:1   & 3200:1 \\
\hline
\end{tabular}

\vskip 0.3cm
{Table 2. Results from Monte Carlo simulation: Efficiency vs background \\
rejection power for different flavors.}
\end{center}
\end{table}

\section{Summary}

In summary, the water \C calorimeter is a cheap and effective detector for 
$\nu$ factories and $\nu$ beams. The performance is excellent
for $\nu_e$ and $\nu_{\tau}$ appearance and $\nu_{\mu}$ 
disappearance from a Monte Carlo simulation.
Such a detector is also very desirable for cosmic-ray physics and astrophysics. 
There are no major technical difficulties although R\&D and detector optimization 
are needed. 

\section*{Acknowledgments}
I would like to thank G. Gratta, S. Wojcicki, L. Wai and H.S. Chen for many 
useful discussions.

\end{document}